# Positive Work Practices. Opportunities and Challenges in Designing Meaningful Work-related Technology

**Matthias Laschke**
University of Siegen
57072 Siegen, Germany
matthias.laschke@uni-siegen.de

**Alarith Uhde**
University of Siegen
57072 Siegen, Germany
alarith.uhde@uni-siegen.de

**Marc Hassenzahl**
University of Siegen
57072 Siegen, Germany
marc.hassenzahl@uni-siegen.de

## Abstract
Work is a rich source of meaning. However, beyond organizational changes, most approaches in the research field of *Meaningful Work* neglected the power of work-related technology to increase meaning. Using two cases as examples, this paper proposes a wellbeing-driven approach to the design of work-related technology. Despite the positive results of our cases, we argue that the use of technology as a means of increasing meaning in the workplace is still in its infancy.

## Author Keywords
Wellbeing-driven design; job design; technology at work; practice-based.

## Introduction
> "Work plays a powerful role in how people understand their lives, the world around them, and the unique niche they fulfill" ([12], p.131).

It is nothing new that most workers not only work for sustenance but also meaning. Besides the possibility of earning a living, Steger and Dik [12] show compellingly that the workplace is a space to demonstrate skills, meet colleagues, and contribute to a higher-level

**Matthias Laschke** has a background in industrial design and holds a doctorate in philosophy with a focus in Human-Computer Interaction. He focuses on the design and aesthetic of transformational objects (i.e., "Pleasurable Troublemakers"), work-related technology, and autonomous systems. His work has been published and discussed in various national and international books and magazines such as the New York Time, Wired, Fast Company, and the R&D Salon of the Museum of Modern Art, New York.

process. The broad research field of *Meaningful Work* comprises of approaches that inquire how employees "find meaning in work" and "approach, enact, and experience their work and workplaces" ([10], p.92). Throughout the years, several models for good and meaningful work have been proposed. Especially approaches in the subfield of Job Design try to understand and represent frameworks and psychological theories on how the work environment and the intrinsic motivation of workers contribute to meaningful work. Beyond early movements such as *Scientific Management* by Taylor, which looked at the design of the work environment from a functional and predominantly economic perspective, more recent approaches began to focus explicitly on the motivation and wellbeing of workers. The *Job Characteristics Model* [8] (JCM) by Hack-man and Oldham, Herzberg's *Two-Factor Theory* [4], or theories by Porter and Lawler [9] are some of the most prominent examples that started to consider the experiences and motivation of workers as crucial aspects in job design. Contemporary approaches such as job crafting [14] not only include the experience and motivation of employees, but they actively involve employees to develop their own work into something more fulfilling.

What most models have in common is that they address the *intrinsic motivation* of employees by making use of theories such as *Self Determination Theory* [1] and the satisfaction of basic needs (e.g., autonomy, relatedness, or competence) as an essential mechanism. However, for many reasons, most work motivation theories neglect the role of work-related technology in work environments. Motivation and meaning are primarily addressed through non-technical organizational changes such as new regulations, work processes, or individual training. Although work-related technology always played a role in productivity [13], its effect on wellbeing was neglected. We argue that integrating motivating work practices in work-related technologies would increase employees' wellbeing. Moreover, work-related technologies are already part of existing workplaces and practices and do not necessarily require complex reorganization of work. At this point, Human-Computer Interaction (HCI) and UX research could provide a worthwhile contribution.

Given the potential of work-related technology to change work practices, we believe that job satisfaction and wellbeing should be actively addressed through design. Therefore, we propose a wellbeing-driven approach to the design of work-related technology that focuses on positive work-practices. In the following, we will briefly describe our approach and present two case studies. Subsequently, we will then discuss challenges that emerged during the cases

**Wellbeing-Driven Design of Work-related Technology**
In HCI, several approaches provide a theoretical foundation on how interactive technology can explicitly address and improve subjective wellbeing and motivation (e.g., [2,3,7,15]).

Based on Hassenzahl's approach of Experience Design [3], we [5,6] focus on the fulfillment of psychological needs, such as competence, relatedness, or popularity through technology use. However, while needs offer guidance, they remain quite abstract. *Social Practices* [11] offer a reasonable lens to look at the work-place and to better understand how needs are fulfilled through interacting with technology. The elements of

**Alarith Uhde** is a doctoral student in Human-Computer Interaction at Siegen University. His research is focused on individual and social consequences of technology use. For the past three years, he worked in a project on technology-enhanced, autonomous shift scheduling in healthcare institutions, empowering nurses to reclaim control of their working hours and solve planning conflicts in a fair way, without the need of a supervisor. Alarith has a background in cognitive psychology and memory research.

*Social Practices* described by Shove et al. [11] are *Materials* (objects, tools, and infrastructures), *Competence* (knowledge and embodied skills), and *Meanings* (cultural conventions, expectations, and socially shared meanings) provide on the one hand a structure to collect and understand existing work practices. On the other hand, they indicate ways to redesign them (e.g., by changing the material – i.e., work-related technology). Hence, our wellbeing-oriented process includes a first step of gathering successful (i.e., meaningful) work practices and take them as inspiration to the (re)design of work-related technology in the second step, evoking positive work practices through the interaction and functionalities offered.

In the following, we briefly describe two case studies from two different work domains that applied the outlined approach.

## Case 1. Improving Radiologists' Wellbeing Through Medical Technology

The case [6] was part of a research collaboration commissioned by one of the world's leading providers of medical technology (MTP) that develops radiological equipment, such as CT scanners, magnetic resonance imaging (MRI) scanners, as well as imaging software. Its main business is to sell technology to healthcare providers (HP) that offer radiological imaging (i.e., using technology) and diagnostics (i.e., done by employed radiologists) to referring physicians. A high workload characterizes the workplace of radiologists due to rising case numbers (also as a result of increasingly efficient technology) and cost pressure (due to competition).

Nevertheless, radiologists are highly sought-after professionals. To increase staff retention, the quality of work, and job satisfaction are very important for HP. The main question of the project was whether a wellbeing-driven approach to design technology could innovate MTP's products and business through an increased wellbeing and job satisfaction of radiologists that work with their products. It should be noted that MTP was skeptical as to whether the approach could innovate their products or whether wellbeing would play a role at all for their customers (HP) and users (radiologists).

In sum, against all skepticism, we found several informal work practices that had the potential to increase radiologists' subjective wellbeing and the HP's business (i.e., for the benefit of the MTP's business). For instance, radiologists record interesting and typical (i.e., pathognomonic) cases in paper notebooks or Excel charts because they have personal significance. Additionally, they also try to get feedback from referrers, for example by calling them. Throughout the case, we used several informal work practices as inspiration to design and prototype two software applications that address the radiologists' wellbeing and job satisfaction. A brief evaluation of the two applications showed that they potentially increase the wellbeing of radiologists and improve the business of the HP and MTP.

In addition to the generally positive results, the case revealed some interesting findings. Initially, it was marked by many breaks and skepticism on the part of the MTP. It might be that subjective wellbeing, and positive practices and their business benefits seem vague and intangible compared to the more established requirements, especially if the success of a business is based on technological progress. In contrast, most of

**Marc Hassenzahl** is Professor for "Ubiquitous Design / Experience and Interaction" at the University of Siegen. He is interested in the positive affective and motivational aspects of interactive technologies – in short: User Experience. He is author of "Experience Design. Technology for all the right reasons." With his group of designers and psychologists, he explores the theory and practice of designing pleasurable, meaningful and transforming interactive technologies.

the findings did not seem new to the HP and radiologists. However, none of them were part of any formal work practice or process. To them, it did not seem appropriate to focus on the wellbeing and job satisfaction of radiologists, although their importance and business benefits were known.

## Case 2. Computer-supported Self-scheduling for Healthcare Workers

The case was part of a government-funded research project whose goal was the development of a tool for computer-supported shift planning for healthcare workers. Usually, shift schedules are created manually. In the future, nurses should be able to make the planning themselves with the support of computers. The software in this area primarily focuses on the effectiveness of the process and compliance with legal regulations, while the motivation and subjective wellbeing of nurses whose work and private lives are planned play a subordinate role. On the current job market, especially nurses are highly sought-after professionals. Thus, job satisfaction is important for any service provider in this domain in the competition for new employees. It should again be mentioned that a manager of the funding initiative was skeptical as to whether the subjective wellbeing and motivation of employees plays a role at all in shift scheduling. After all, in his view, shift planning is only about the organization of work that needs to be done.

Luckily (or, of course), the case showed that many different practices exist around shift planning that potentially increase the wellbeing of nurses. For example, nurses note requests for shifts where they would preferably not be scheduled to do things in their private lives (e.g., concerts or visits to the doctor). In addition, they fill in for each other and swap shifts to individually balance work and private life. These and other practices contribute to the wellbeing of the nurses and a positive team spirit, but are not addressed by rostering software. The evaluation of our prototypical system showed that the well-being of nurses and their participation in the planning process could be increased.

Beyond the positive results, the case also revealed some further insights. For instance, although computer-assisted systems could complete shift schedules much earlier than a manual planner, the employer want to publish them as late as possible in order to remain flexible in terms of personnel resources (i.e., to reduce costs). It seems that personnel costs and profits, in contrast to the well-being of workers, have a higher or at least more tangible price.

## Conclusion

The field of meaningful work is a well-researched area. A large variety of approaches show how the motivation and wellbeing of workers leads to happy employees and efficient work. However, the field neglected the role of work-related technology as a means to increase wellbeing. We suggest a wellbeing-driven approach to the design of work-related technology that evokes positive work practices through their functionalities offered. However, although the case studies presented indicate that our approach leads to positive results, we believe that the idea that technology has to be not only perfectly adapted to the work task at hand but should also be explicitly designed to increase meaning. This view is still in its infancy in HCI.